\documentclass[aps,twocolumn,pra,superscriptaddress,amsmath,showpacs,tightenlines]{revtex4}
\usepackage{epsfig,graphicx,times}
\usepackage{amstext}
\usepackage{amsmath}            %serve per le subequazioni
\usepackage{amssymb}            %serve per il simbolo "marchio registrato", \circledR
\usepackage{graphicx}           %serve per le figure eps, ps etcyy
\usepackage{latexsym}
\usepackage{bm}
\begin{document}
%\title{Synthesizing arbitrary phonon states in an optomechanical system}

\title{Engineering of nonclassical motional states in optomechanical systems}

\author{Xun-Wei Xu}
\affiliation{Institute of Microelectronics, Tsinghua University,
Beijing 100084, China}

\author{Hui Wang}
\affiliation{Institute of Microelectronics, Tsinghua University,
Beijing 100084, China}

\author{Jing Zhang}
\affiliation{Department of Automation, Tsinghua University, Beijing
100084, P. R. China}
\affiliation{Tsinghua National Laboratory for
Information Science and Technology (TNList), Tsinghua University,
Beijing 100084, China}

\author{Yu-xi Liu}\email{yuxiliu@mail.tsinghua.edu.cn}
\affiliation{Institute of Microelectronics, Tsinghua University,
Beijing 100084, China} \affiliation{Tsinghua National Laboratory for
Information Science and Technology (TNList), Tsinghua University,
Beijing 100084, China}

\date{\today}

\begin{abstract}
We propose to synthesize arbitrary nonclassical motional states in optomechanical systems by using sideband excitations and photon blockade. We first demonstrate that the Hamiltonian of the optomechanical systems can be reduced, in the strong single-photon optomechanical coupling regime when the photon blockade occurs, to one describing the interaction between a driven two-level trapped ion and the vibrating modes, and then show a method to generate target states by using a series of classical pulses with desired frequencies, phases, and durations. We further analyze the effect of the photon leakage, due to small anharmonicity, on the fidelity of the expected motional state, and study environment induced decoherence. Moreover, we also discuss the experimental feasibility and provide operational parameters using the possible experimental data.

\end{abstract}

\maketitle \pagenumbering{arabic}

\section{Introduction}
Whether macroscopic mechanical resonators
behave quantum mechanics is a long-outstanding debate of the
fundamental physics~\cite{physrep,review1,review2}. Recent
experimental progresses on, e.g., ground-state cooling and the
fabrication of high-frequency mechanical resonators, push forward
the process to end this debate. In existing literatures, several
methods have been proposed to cool the mechanical resonators to their
ground state in various types of the nano-structures, e.g., doubly
clamped beams, singly clamped cantilevers, radial breathing modes of
micro-toroids, and membranes. The potential applications of
mechanical resonators in the quantum regime can be referred to,
e.g., quantum information processing and sensitive quantum detection
of very weak forces.

It is well known that quantum superpositions are main resources
for quantum information processing. Many theoretical proposals and
experimental demonstrations have been presented to generate and
manipulate quantum superposed states. For example, we have
theoretically studied how to generate superpositions of different
Fock states for microwave photons~\cite{liu2004}, and later on
experimentalists produced Fock states~\cite{martinis-nature1} and
arbitrary superpositions~\cite{martinis-nature2} of different Fock
states by coupling a single-mode microwave cavity field to a
superconducting phase qubit. Similarly, particular non-classical
phonon states of the vibrational mode of trapped ions have been
theoretically studied~\cite{Gardiner97,zheng,wei,RMP} and experimentally
demonstrated~\cite{ions1,ions2}. However, the generation of
arbitrary nonclassical motional states (hereafter, we call them as
phonon states) in macroscopic mechanical resonators with
low-frequencies is still an open question.

Macroscopic mechanical resonators in the quantum regime can be
manipulated by integrating them with other quantum components. For
instance, the superpositions of macroscopically distinct quantum
states have been theoretically studied in a mechanical resonator by
coupling it to a charge qubit~\cite{armour}. The quantum ground
state and single-phonon control~\cite{cleland and martinis} have
been experimentally demonstrated for a microwave-frequency
mechanical resonator coupled to a phase qubit. This circuit-QED-like
system~\cite{cleland and martinis} makes it possible to engineer
arbitrary phonon states in a deterministic way as for microwave
photon states~\cite{liu2004,martinis-nature1,martinis-nature2}. The recent studies demonstrate
that optomechanical systems~\cite{review1,review2} can provide another platform to
control and manipulate the quantum states of the low-frequency
mechanical resonator by coupling it to a cavity field. In particular, experiments~\cite{strong1,strong2,strong3,strong4} showed
that the optomechanical systems are approaching the strong single-photon
coupling regime.

It has been shown that the photon blockade can
occur~\cite{Imamoglu,blockade1,blockade2,binghe,JQLiao1,JQLiao2} in the strong single-photon optomechanical coupling when single-photons pass
through the cavity of the optomechanical system. We here study a method to
synthesize arbitrary nonclassical phonon states in optomechanical
systems by using photon blockade and a series of sideband excitations with desired durations. We mention that the red
sideband excitations were studied theoretically~\cite{theory1,theory2}
and experimentally
~\cite{sideband1,sideband2,sideband3,sideband4,Painter1,Painter-new}
for the ground state cooling of the mechanical resonators. In
contrast to the method of the
measurement-based~\cite{non1,Pepper,kim} non-Gaussian phonon state
generation~\cite{non2,non3} in optomechanical systems, our method is
deterministic one as for microwave single-phonon
generation~\cite{cleland and martinis}. But here we need sideband excitations
to make the low frequency mechanical resonator to resonantly interact with the
high frequency cavity field, assisted by the driving field with the frequency matching condition, the microwave single-phonon
generation requires no sideband excitations~\cite{cleland and martinis}.

The purpose of this paper is to present a method on the preparation of the arbitrary nonclassical phonon states in optomechanical systems. We will mainly analyze detailed steps, possible errors and experimental feasibilities. In Sec.~II, the theoretical model of the optomechanical system is introduced, an effective Hamiltonian is derived in the strong single-photon optomechanical coupling
regime. We find that this effective Hamiltonian is equivalent to one of trapped ions~\cite{BlockleyEPL}. In Sec.~III, we show how to synthesize phonon states by using sideband excitation and the effective Hamiltonian derived in Sec.~II. In Sec.~IV, we analyze the effect of the photon leakage on the fidelity of the expected target state due to small anharmonicity. In Sec.~V, the environmental effect on prepared states is further studied. Moreover, we discuss the experimental feasibility and provide operational parameters in Sec.~VI. The conclusions are finally given in Sec.~VII.

\section{Theoretical model}

\begin{figure}
\includegraphics[bb=55 205 590 420, width=8.5 cm, clip]{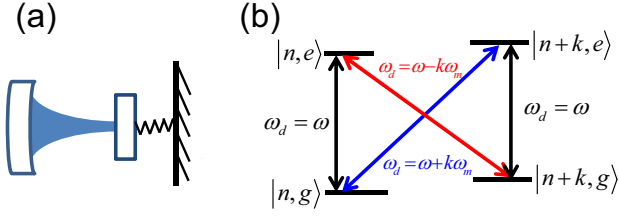}
%\centerline{\epsfxsize=8.5cm\epsfbox{fig1.eps}}
\caption[]{(Color online) (a) Schematic diagram for optomechanical
systems with the radiation-pressure type interaction: the cavity can
be in either the optical, or the microwave, or the radio-wave
regime; and the mechanical resonator can be doubly clamped beams,
singly clamped cantilevers, radial breathing modes of micro-toroids,
and membranes. (b) Three different transition processes are
presented by the black (carrier), red ($k$-phonon red sideband
excitation) and blue ($k$-phonon blue sideband excitation) arrow
lines.}\label{fig1}
\end{figure}

We study an optomechanical system,
which can be the membrane coupled to an optical cavity, or the
optical cavity with one-end movable mirror, or the superconducting
transmission line resonator coupled to a mechanical beam. As
schematically shown in Fig.~\ref{fig1}(a), such system has the
radiation-pressure-type interaction, and the Hamiltonian of the
system driven by a classical field can be written as
\begin{eqnarray}\label{eq:1}
H&=&\hbar\omega_{c} a^{\dagger}a +\hbar\omega_{m}b^{\dagger}b+\hbar g
a^{\dagger}a(b^{\dagger}+b) \nonumber \\
&&+\hbar \Omega
\left[a^{\dagger}e^{-i(\omega_{d}t+\phi_{d})}+\text{h.c.}\right].
\end{eqnarray}
Here, $a^{\dagger}(a)$ is the creation (annihilation) operator of
the cavity field with the frequency $\omega_{c}$, and
$b^{\dagger}(b)$ is the creation (annihilation) operator of the
mechanical resonator with the frequency $\omega_{m}$. The parameter
$g$ describes the coupling strength between the cavity field and the
mechanical resonator. The parameter $\Omega$ is the coupling
strength between the cavity field and the external driving field
with the frequency $\omega_{d}$ and the phase $\phi_{d}$.

If an unitary transform
$U=\exp[ga^{\dagger}a(b^{\dagger}-b)/\omega_{m}]$ is applied to
Eq.~(\ref{eq:1}), then the Hamiltonian in Eq.~(\ref{eq:1}) becomes
\begin{eqnarray}\label{eq:s2}
H_{\rm eff} &=&\hbar \omega a^{\dag }a-\hbar \frac{%
g^{2}}{\omega _{m}}a^{\dag }a^{\dag }aa +\hbar \omega _{m}b^{\dag }b\nonumber\\
&&+\hbar \Omega \left\{ a^{\dag }e^{\left[ \eta \left( b^{\dag
}-b\right) -i\left( \omega _{d}t+\phi _{d}\right) \right]
}+\text{h.c.}\right\},
\end{eqnarray}
where $\omega=\omega_{c}-g^{2}/\omega_{m}$ and $\eta=g/\omega_{m}$. It is obvious that the
energy structure of the photon Hamiltonian, corresponding to the first two terms in
the right hand of Eq.~(\ref{eq:s2}), becomes anharmonic one due
to the photon-photon interaction induced by the radiation pressure.
Moreover, the nonlinear photon-photon interaction
term $\hbar g^2 a^{\dagger 2}a^2/\omega_{m}$ guarantees the photon
blockade~\cite{blockade1} in the optomechanical systems with the
strong coupling strength $g$ and low dissipation of the cavity
field, i.e., $(g^2/\omega_{m})>\gamma_{c}$ with the decay rate
$\gamma_{c}$ of the cavity field. In this case, the driving field
couples only two lowest energy levels $|0\rangle$ and $|1\rangle$ of
the cavity field, and Eq.~(\ref{eq:s2}) can be further reduced to
\begin{eqnarray}\label{eq:2}
H_{\rm tw}=\hbar\frac{\omega}{2}\sigma_{z}+\hbar\omega_{m}b^{\dagger}b
+\hbar\left\{\Omega(t)\sigma_{+}
e^{\eta(b^\dagger-b)}+\text{h.c.}\right\},
\end{eqnarray}
under the two-level approximation for the cavity field with
\begin{equation*}
\Omega(t)=\Omega \exp{\left[-i(\omega_{d}t+\phi_{d})\right]}.
\end{equation*}
Here, we redefine the photon operators $a^{\dagger}$ and $a$ via the ladder
operator $\sigma_{+}=|1\rangle\langle 0|$ and
$\sigma_{-}=|0\rangle\langle 1|$ in the basis of two states $|0\rangle$ and $|1\rangle$ of the cavity field. We also define the Pauli operator
$\sigma_{z}=|1\rangle\langle 1|-|0\rangle\langle 0|$. Hereafter, we use $|e\rangle$ and
$|g\rangle$ to denote the single-photon excited state $|1\rangle$
and the vacuum (ground) state  $|0\rangle$  of the cavity field,
respectively, i.e., $|1\rangle\equiv |e\rangle$ and $|0\rangle\equiv
|g\rangle$. Moreover, the states $|k\rangle$ with $ k=1,2,\cdots,N$
denote the phonon number states of the mechanical resonator with
the transform $U$.

The effective Hamiltonian in Eq.~(\ref{eq:2}) is similar to one
that describes the interaction between a classical driving field and
a single two-level trapped cold ion vibrating along
one-direction~\cite{BlockleyEPL}. That is, the two-level trapped
cold ion, the vibrating mode of the trapped ion  and the classical
driving field are equivalent to the two-level system constructed by
two lowest energy levels of the cavity field, the vibrating mode of
the mechanical resonator, the classical field applied to the cavity,
respectively. The parameter $\eta$ is equivalent
to the Lamb-Dicke parameter in the system of trapped ion.
The third term in Eq.~(\ref{eq:2}) can
be further written as
\begin{eqnarray}\label{eq:3}
H_{\rm th}=\hbar\Omega(t)\,
e^{-\frac{\eta^2}{2}}\sigma_{+}\sum_{j,l}\frac{(-1)^l\eta^{(j+l)}b^{\dagger
j}b^{l}}{j!l!}+{\rm h.c.},
\end{eqnarray}
which can be reduced to the carrier, or red sideband excitation or blue sideband excitation
precess in different resonant conditions with the language of trap
ions~\cite{BlockleyEPL}.

As schematically shown in Fig.~\ref{fig1}, the $k$-phonon red sideband excitation process links the transitions
between $|n,e\rangle$ and $|n+k,g\rangle$ under the condition
$\omega_{d}=\omega-k\omega_{m}$, with an effective Rabi frequency
\begin{equation}\label{eq:R5}
\Omega_{n,k}=\Omega e^{(-\eta^2/2)}
\eta^{k}\sqrt{\frac{(n+k)!}{n!}}\sum_{j=0}^{n}\frac{(-1)^j\eta^{2j} C_{n}^{j}}{(j+k)!}
\end{equation}
here $C_{n}^{j}=n!/[j!(n-j)!]$, i.e., the cavity field transits from the ground
(excited) state to the excited (ground) state by absorbing
(emitting) $k$ phonons assisted by the external field. The
$k$-phonon blue sideband excitation links transition between
$|n,g\rangle$ and $|n+k,e\rangle$ under the condition
$\omega_{d}=\omega+k\omega_{m}$, with an effective Rabi frequency
$\Omega_{n,k}$ as shown in Eq.~(\ref{eq:R5}), i.e., the cavity field transits from the ground
(excited) state to the excited (ground) state by emitting
(absorbing) $k$ phonons with the help of the external field.
The carrier process links the transitions between $|n,e\rangle$ and $|n,g\rangle$ under the
condition $\omega_{d}=\omega$, with an effective Rabi frequency $\Omega_{n,0}$ given by Eq.~(\ref{eq:R5}) with $k=0$. Therefore, no phonon absorption or
emission occurs and the external field only flips the photon states in the carrier process. Under the condition (Lamb-Dicke limit)
\begin{equation}
\eta\sqrt{\overline{n}+1}=\frac{g}{\omega_{m}}\sqrt{\overline{n}+1}\ll 1
\end{equation}
with the average phonon number $\overline{n}$ of the mechanical
vibration, we have
\begin{equation}
\exp{[\eta(b^\dagger-b)]}\approx
1+\eta(b^\dagger-b).
\end{equation}
In this case, only a single-phonon transition
occurs with the help of the driving field for $k=1$ in Fig.~\ref{fig1}(b).
The time evolution operators for the carrier, red and blue
sideband precesses, described by $U^{c}_{0}(t^{c})$,
$U^{r}_{n,k}(t^{r})$, and $U^{b}_{n,k}(t^b)$,
are given in Appendix~\ref{Evolution}, here the superscripts
denote different processes, e.g., $r$ denotes the red sideband excitation process.

\section{Synthesizing phonon states}
We have reduced the Hamiltonian in Eq.~(\ref{eq:1}) of the driven optomechanical
system to Eq.~(\ref{eq:3}), which is similar to that of
trapped ions~\cite{BlockleyEPL}. Thus, arbitrary superpositions of different phonon number states $|k\rangle$
\begin{equation}\label{eq:s12}
\left\vert \psi \right\rangle =\overset{N}{\underset{k=0}{\sum }}%
c_{k}\left\vert k\right\rangle \text{, \ \ \ }\overset{N}{\underset{k=0}{%
\sum }}\left\vert c_{k}\right\vert ^{2}=1.
\end{equation}%
can be generated by using similar method as in the system of the
trap ions~\cite{Gardiner97,zheng,wei}, where $\left\vert c_{k}\right\vert ^{2}$ is the probability corresponding
to the phonon number state $|k\rangle$. We clarify Eq.~(\ref{eq:s12}) denotes the phonon state without the transform $U=\exp[ga^{\dagger}a(b^{\dagger}-b)/\omega_{m}]$. In the following, our study on state preparation is in the basis with the transform $U$.  However, the target state of the whole system is $|\psi\rangle|g\rangle$, which is not changed with an inversion of the transform $U$ because $U^{\dagger}|g\rangle=U|g\rangle=|g\rangle$.

Under the condition $\eta\ll 1$,  we only consider single-phonon transitions
assisted by the driving field. In this case, the arbitrary phonon state as in Eq.~(\ref{eq:s12}) can be prepared by using the method given in Ref.~\cite{Gardiner97}. We need $N$-step red sideband excitations and $N$-step carrier processes for such state preparation. The whole process can be described as
\begin{equation}
U_{T}(t)\left\vert 0,g\right\rangle = \sum^{N}_{k=0} c_{k} \left\vert k,g\right\rangle\equiv \left[ \sum^{N}_{k=0} c_{k} \left\vert k\right\rangle\right]|g\rangle,
\end{equation}
with a total time evolution operator $U_{T}(t)$, decomposed as
\begin{eqnarray}
U_{T}(t) &=&  U_{1}^{r}\left( t^{r}_{2N}\right)U^{c}\left( t^{c}_{2N-1}\right)\cdots U_{1}^{r}\left( t^{r}_{2(i+1)}\right)U^{c}\left( t^{c}_{2i+1}\right) \nonumber\\
&& \cdots U_{1}^{r}\left( t^{r}_{2}\right)U^{c}\left( t^{c}_{1}\right),
\end{eqnarray}
and total time
\begin{equation}
t = \sum ^{N}_{i=1} \left(  t^{r}_{2i} + t^{c}_{2i-1} \right),
\end{equation}
here the superscript ``$r$" and ``$c$" demote the red-side excitation and carrier process, respectively.
The time intervals and phases in each process can be calculated by following the method given in Refs.~\cite{Gardiner97,Law96,Law98}. The main steps are to find $U^{\dagger}(t)$ such that
\begin{equation}
\left\vert 0,g\right\rangle =U^{\dagger}(t) \sum^{N}_{k=0} c_{k} \left\vert k,g\right\rangle.
\end{equation}
In each step, the conditions
\begin{eqnarray}
&&\left\langle i,g\right\vert  U_{1}^{r\dag}\left( t^{r}_{2i}\right) \left\vert \Psi_{i} \right\rangle = 0,  \label{eq:c11}\\
&&\left\langle i-1,e\right\vert  U^{c\dag}\left( t^{c}_{2i-1}\right) U_{1}^{r\dag}\left( t^{r}_{2i}\right) \left\vert \Psi_{i} \right\rangle = 0, \label{eq:c12}
\end{eqnarray}
should be satisfied. Where
\begin{eqnarray}
\left\vert \Psi_{i} \right\rangle &=& U^{c\dag}\left( t^{c}_{2i+1}\right)U_{1}^{r\dag}\left( t^{r}_{2(i+1)}\right)  \nonumber\\
&& \cdots U^{c\dag}\left( t^{c}_{2N-1}\right) U_{1}^{r \dag}\left( t^{r}_{2N}\right) \sum^{N}_{k=0} c_{k} \left\vert k,g\right\rangle .
\end{eqnarray}
The conditions (\ref{eq:c11}) and (\ref{eq:c12}) provide the equations to determine the time intervals and phases in each process as follow:
\begin{eqnarray}
\tan \left( \Omega _{i,1}t_{2i}^{r}\right) &=& \frac{-i\beta _{i}e^{-i\phi
_{r}^{2i}}}{\alpha _{i}} , \label{eq:16}\\
\tan \left( \Omega _{i-1,0}t_{2i-1}^{c}\right) &=& \frac{i\upsilon _{i}e^{i\phi
_{c}^{2i-1}}}{\mu _{i}}
\end{eqnarray}
where
\begin{eqnarray}
&&\alpha_{i}= \left\langle i-1,e | \Psi_{i} \right\rangle ,  \\
&&\beta_{i}= \left\langle i,g | \Psi_{i} \right\rangle ,  \\
&&\mu_{i}= \left\langle i-1,g \right\vert  U_{1}^{r\dag}\left( t^{r}_{2i}\right) \left\vert \Psi_{i} \right\rangle ,  \\
&&\nu_{i}= \left\langle i-1,e \right\vert  U_{1}^{r\dag}\left( t^{r}_{2i}\right) \left\vert \Psi_{i} \right\rangle .\label{eq:21}
\end{eqnarray}
Therefore, using Eqs.~(\ref{eq:16}-\ref{eq:21}), we can obtain all time intervals and phases, and thus the expected state can be prepared.

Out of the regime $\eta\ll 1$, the arbitrary phonon state in Eq.~(\ref{eq:s12}) can be prepared by sequentially applying $N$
red-sideband excitations after a carrier process as shown in Ref.~\cite{wei}. That is, the
cavity field is first driven by the classical field with the frequency
matching condition $\omega_{d}=\omega$, then with a time interval
$t^{c}_{0}$, the system evolves to
\begin{equation}
\left\vert 0,g\right\rangle \overset{U_{0,0}^{c}\left( t^{c}_{0}\right) }{%
\rightarrow }c_{0}\left\vert 0,g\right\rangle -ie^{-i\phi
_{c}^{0}}\sin \left( \Omega _{0,0}t^{c}_{0}\right) \left\vert
0,e\right\rangle
\end{equation}
according to the evolution operator $U_{0,0}^{c}\left( t^{c}_{0}\right)$
of the carrier process in Eq.~(\ref{eq:s11}) of the Appendix~\ref{Evolution}, here  $c_{0}=\cos
\left( \Omega _{0,0}t^{c}_{0}\right)$. After the carrier process, $N$
red-sideband excitations are sequentially applied to the cavity with
the frequency matching conditions
$\omega_{d}=\omega-\omega_{m},\,\,\omega-2\omega_{m},\,\,\cdots,\,
\omega-N\omega_{m}$ for the time intervals $t^{r}_{1},\,\cdots,\,t^{r}_{N}$,
respectively. Then the system will evolve according to the evolution
operators in Eq.~(\ref{eq:s8})  of the Appendix~\ref{Evolution} and the target state can be obtained.
For example, in the first red-sideband excitation with the evolution
operator $U_{0,1}^{r}\left( t^{r}_{1}\right)$, if the phase and the time
interval are chosen such that
\begin{equation}
c_{1}=e^{i\left( \phi _{r}^{1}-\phi
_{c}^{0}\right) }\sin \left( \Omega _{0,0}t^{c}_{0}\right) \sin \left(
\Omega _{0,1}t^{r}_{1}\right),
\end{equation}
then state of the system evolves to
\begin{eqnarray}
\left\vert \Psi \right\rangle &=& c_{0}\left\vert 0,g\right\rangle +c_{1}\left\vert 1,g\right\rangle \nonumber\\
&&-ie^{-i\phi _{c}^{0}}\sin \left( \Omega _{0,0}t^{c}_{0}\right) \cos
\left( \Omega _{0,1}t^{r}_{1}\right) \left\vert 0,e\right\rangle.
\end{eqnarray}
after the first red sideband excitation. We can properly chose
the the phase $\phi_{r}^{k}$ and duration $t^{r}_{k}$ of the $N$ red
sideband excitations such that
\begin{widetext}
\begin{equation}
c_{k}=\left\{
\begin{array}{ll}
\cos \left( \Omega _{0,0}t^{c}_{0}\right)  & k=0, \\
\left( -1\right) ^{k-1}e^{i\left( \phi _{r}^{k}-\phi _{c}^{0}\right)
}\sin \left( \Omega _{0,0}t^{c}_{0}\right)
\overset{k-1}{\underset{j=1}{\prod }}\cos \left( \Omega
_{0,j}t^{r}_{j}\right) \sin \left( \Omega _{0,k}t^{r}_{k}\right)  &
1\leqslant k\leqslant N-1 ,\\
\left( -1\right) ^{N-1}e^{i\left( \phi _{r}^{N}-\phi _{c}^{0}\right)
}\sin \left( \Omega _{0,0}t^{c}_{0}\right)
\overset{N-1}{\underset{j=1}{\prod }}\cos
\left( \Omega _{0,j}t^{r}_{j}\right)  & k=N,
\end{array}%
\right.
\end{equation}
\end{widetext}
then we can obtain
\begin{equation}
\left\vert \Psi \right\rangle= \underset{k=0}{\overset{N}{\sum
}}c_{k}\left\vert k,g\right\rangle \equiv
\underset{k=0}{\overset{N}{\sum }}c_{k}\left\vert
k\right\rangle\otimes\vert g\rangle,
\end{equation}%
which is a product state of the target phonon state
in Eq.~(\ref{eq:s12}) and the ground state  $\left\vert
g\right\rangle $ of the cavity field.

\section{Information leakage due to small anharmonicity}

\begin{figure}
\includegraphics[bb=44 245 330 400, width=8 cm, clip]{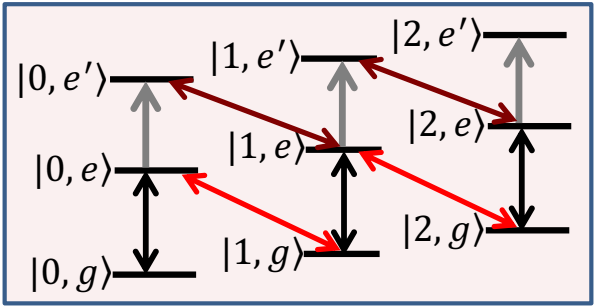}
\caption[]{(Color online) Schematic diagram for the information
leakage to the third level due to the small arharmonicity. Here, two
lowest horizontal lines in each column linked by the vertical black
line with two arrows denote two-level approximation with the carrier
process. The gray vertical line with the arrow pointed to the top
line in each column simply denotes the information leakage in the
carrier process. However, each red (dark red) slanted line with two
arrows pointed to two black lines in different columns denotes the
red sideband excitation (information leakage in the red sideband
excitation). The first and second letter in the states, e.g, $|g,0\rangle$, denote that the
cavity field  and the mechanical resonator are the ground state $|g\rangle$ and the vacuum state $|0\rangle$, respectively.}\label{fig2}
\end{figure}

In the above, our discussions for generating an arbitrary phonon state
are based on the two-level approximation of the cavity
field. That is,  the photon states is confined to the two-dimensional Hilbert space in the basis of
photon states $|g\rangle$ and $|e\rangle$ (or $|0\rangle$ and $|1\rangle$). However, we know that the anharmonicity of the cavity field
induced by the radiation pressure is not very large because the
optomechanical interaction is usually not very strong.
Therefore, the fidelity of the prepared target phonon state
will be affected by the upper levels of the cavity field. To study how the
small anharmonicity of the cavity field affects the fidelity of prepared nonclassical
phonon states, we now study, as an example, the interaction between
the mechanical resonator and three-level photon system, formed by three
lowest energy levels $|0\rangle\equiv |g\rangle$, $|1\rangle\equiv
|e\rangle$ and $|2\rangle\equiv |e^{\prime}\rangle$ of the cavity
field. The transition frequency between the states $|e\rangle$ and
$|e^{\prime}\rangle$ is assumed as $\omega+\delta$. The parameter
$\delta$ characterizes the anharmoncity of the energy levels of the
cavity field. The harmonic and the two-level model can be recovered
when $\delta=0$ and $\delta=\infty$, respectively. In optomechanical
systems, the anharmonicity is $\delta=-2 g^2/\omega_{m}$, which is a
negative number, i.e., the transition frequency between the states
$|e\rangle$ and $|e^{\prime}\rangle$ is smaller than that between
the states $|g\rangle$ and $|e\rangle$. As schematically shown in
Fig.~\ref{fig2}, the Hamiltonian $H_{\rm thr}$ between the
three-level photon system and the mechanical mode can be written as
\begin{eqnarray}
H_{\mathrm{thr}} &=&\hbar \omega _{m}b^{\dagger }b+\hbar \omega
|e\rangle \langle e|+\hbar (2\omega +\delta )|e^{\prime }\rangle
\langle e^{\prime }|\nonumber \\
&+&\hbar \left\{ \Omega (t)\left[ |e\rangle \langle
g|+\sqrt{2}|e^{\prime }\rangle \langle e|\right] e^{\eta (b^{\dagger
}-b)}+\text{h.c.}\right\},\nonumber\\ \label{eq:s20}
\end{eqnarray}
by projecting the cavity field operators $a^{\dagger}$  and $a$ in Eq.~(\ref{eq:s2}) to three eigenstates
$|g\rangle,\,|e\rangle$ and $|e^{\prime}\rangle$ of the cavity field. Here, the parameter
$\Omega(t)$ in Eq.~(\ref{eq:s20}) is the same as that in
Eq.~(\ref{eq:2}).

If the ratio $\eta$ is big enough, an arbitrary phonon state can be prepared
by several red-sideband excitations after a carrier process, then the information leakage only occurs in the carrier process.
In the carrier process for the time interval $t_{0}^{c}=\pi/(2\Omega_{0,0})$, the cavity field
is prepared to its first excited state $|e\rangle$ from the ground
state $ | g \rangle $ under the two-level approximation in Eq.~(\ref{eq:2}). However,
when the information leakage from the first to the second exited
state is considered, the wavefunction of the cavity field at the time $t$ should
be written as
\begin{equation}
|\varphi(t)\rangle=
c_{g}(t)|g\rangle+c_{e}(t)|e\rangle+c_{e^{\prime}}(t)|e^{\prime}\rangle.
\end{equation}
Three coefficients $c_{g}(t)$, $c_{e}(t)$, and
$c_{e^{\prime}}(t)$ can be obtained by solving the Schrodinger
equation with given initial state $|g\rangle$. Thus,
under the condition $\Omega \ll |\delta|$, the fidelity of preparing
the excited state $|e\rangle$ can be approximately given as
\begin{equation}\label{eq:5}
F=|\langle \varphi(t_{0}^{c})|e\rangle|^2\approx
\left|\left(1-\frac{3\Omega^2}{2\delta^2}\right)\sin\left[\frac{\pi}{2}\left(1-\frac{\Omega^2}{2\delta^2}\right)\right]\right|^2.
\end{equation}
This type of information leakage has been studied in superconducting phase qubit systems~\cite{AminLTP}.

If the ratio
$\eta$ is very small, then we need several carrier processes to generate the arbitrary
superpositions of different phonon states. Thus, the fidelity calculation becomes complicated when the information leakage is
included. Below, we discuss the information leakage in the
limit $\eta \ll 1$. After we neglect the terms of
$O\left(\eta^{2}\right)$, the Hamiltonian in Eq.~(\ref{eq:s20})
becomes
\begin{eqnarray}
\widetilde{H}_{\mathrm{thr}} &=&\hbar \omega _{m}b^{\dagger }b+\hbar
\omega |e\rangle \langle e|+\hbar (2\omega +\delta )|e^{\prime
}\rangle \langle e^{\prime }|
\nonumber \\
&+&\hbar \left\{ \Omega (t)\left[ |e\rangle \langle g|+\sqrt{2%
}|e^{\prime }\rangle \langle e|\right] \left[ 1+\eta b^{\dagger }-\eta b
\right]+\text{h.c.}\right\}.\nonumber\\
\end{eqnarray}%

As an example, we analyze the effect of the third level
$|e^{\prime}\rangle$ of the cavity field
on the fidelities for preparing the phonon states $|2\rangle $ and $
(|0\rangle -|2\rangle )/\sqrt{2}$ from the initial state
$|0,g\rangle$ by using
the carrier and the single-phonon red sideband excitation processes.
Let us first calculate the fidelity for preparing the state
$|2\rangle $ with the following steps:
\begin{eqnarray*}
|0,g\rangle&&\underrightarrow{\text{(i) carrier}}\;\; |0,e\rangle\;\; \underrightarrow{\text{(ii) red sideband excitation
}}\;\;|1,g\rangle\\
&&\underrightarrow{\text{(iii) carrier}}\;\; |1,e\rangle\;\;
\underrightarrow{\text{(iv) red sideband
excitation}}\;\;|2,g\rangle.
\end{eqnarray*}
At the initial time $t_{0}$, we have $c_{0,g}(t_{0})=1$, and the
other coefficients are equal to zero. In the step (i), the system is
excited to the state $|0,e\rangle $ from the state $|0,g\rangle $ by
the external field with the carrier process, there is information
leakage to the state $|0,e^{\prime }\rangle $ with a probability
$|c_{0,e^{\prime}}(t_{1}^{c})|^2$ in this process. For the time interval
$t^{c}_{1}=\pi /(2\Omega) $, when the cavity field is prepared to
the first excited state $|e\rangle$ for the two-level approximation,
the coefficient $\widetilde{c}_{0,e}\left( t_{1}\right)$, that the
system is in the state $|0,e\rangle$ at the time
$t_{1}=t_{0}+t^{c}_{1}$, can be given via Eq.~(\ref{eq:s34}) in the Appendix~\ref{Leakage} as
\begin{equation}\label{eq:s39-1}
\widetilde{c}_{0,e}\left( t_{1}\right) \approx -i f_{ce}
\end{equation}
when the third level of the cavity field is included, where
\begin{equation}\label{eq:s39}
f_{ce}=\left( 1-\frac{3\Omega ^{2}}{2\delta ^{2}}\right) \sin \left[ \frac{%
\pi }{2}\left( 1-\frac{\Omega ^{2}}{2\delta ^{2}}\right) \right].
\end{equation}
In the step (ii), the system evolves to the state $|1,g\rangle $
from the state $|0,e\rangle$ via the red-sideband excitation process
and there is no information leakage in this step. From
Eq.~(\ref{eq:s29}) in the Appendix~\ref{Leakage} with the time interval $t^{r}_{2}=\pi /(2\Omega
\eta) $, the coefficient $\widetilde{c}_{1,g}\left( t_{2}\right)$,
that the system is in the state $|1,g\rangle$ at the time
$t_{2}=t_{1}+t^{r}_{2}$, can be approximately given as
\begin{equation}
\widetilde{c}_{1,g}\left( t_{2}\right) \approx
i\widetilde{c}_{0,e}\left( t_{1}\right).
\end{equation}
In the step (iii), the system is prepared to the
state $|1,e\rangle $ from the state $%
|1,g\rangle $ via the second carrier process with the information
leakage to the state $|1,e^{\prime }\rangle$. With the time interval
$t^{c}_{3}=\pi /(2\Omega) $, the coefficient
$\widetilde{c}_{1,e}\left( t_{3}\right)$, that the system is in the
state $|1,e\rangle$ at the time $t_{3}=t_{2}+t^{c}_{3}$, can be
obtained via Eq.~(\ref{eq:s34}) in the Appendix~\ref{Leakage} as
\begin{equation}
\widetilde{c}_{1,e}\left( t_{3}\right) \approx -i f_{ce} \widetilde{c}_{1,g}\left( t_{2}\right).
\end{equation}
In the step (iv), the system evolves to the state $|2,g\rangle $ via
the second red sideband excitation with the time interval $
t^{r}_{4}=\pi /(2\sqrt{2}\eta \Omega) $, there is information
leakage to the state $|0,e^{\prime}\rangle$ in this step. Using
Eq.~(\ref{eq:s37}) in the Appendix~\ref{Leakage}, the coefficient $\widetilde{c}_{2,g}\left(
t_{4}\right)$, that the system is in the state $|2,g\rangle$ at the
time $t_{4}=t_{3}+t^{r}_{4}$, can be given as
\begin{equation}\label{eq:s42}
\widetilde{c}_{2,g}\left( t_{4}\right) \approx i f_{rg}
\widetilde{c}_{1,e}\left( t_{3}\right).
\end{equation}
where
\begin{equation}\label{eq:s43}
f_{rg}=\left( 1-\frac{2\left( \eta \Omega \right) ^{2}}{\delta
^{2}}\right) \sin \left[ \frac{\pi }{2}\left(
1-\frac{3}{4}\frac{\left( \eta \Omega \right) ^{2}}{\delta
^{2}}\right) \right].
\end{equation}
Because the fidelity to prepare the state $|2\rangle $ is defined as
\begin{equation}\label{eq:s44}
F_{1}=\left\vert \left\langle 2,g | \varphi \left( t_{4}\right) \right\rangle \right\vert
^{2}.
\end{equation}%
which can be given as
\begin{equation}\label{eq:s45}
F_{1}=\left\vert \widetilde{c}_{2,g}\left( t_{4}\right) \right\vert
^{2}.
\end{equation}%
From Eq.~(\ref{eq:s39-1}) to Eq.~(\ref{eq:s43}), we can approximately
obtain
\begin{equation}\label{eq:s46}
F_{1}\approx \left\vert f_{rg}\left( f_{ce}\right)^{2} \right\vert
^{2}.
\end{equation}

Similarly, the preparation of the state $(|0\rangle -|2\rangle
)/\sqrt{2}$ also needs four steps as for that of the state
$|2\rangle$, but with different time intervals.  Thus by using
similar calculation steps, the fidelity $F_{2}$, for preparing the
superposition $(|0\rangle -|2\rangle )/\sqrt{2}$, can be given as
\begin{equation}\label{eq:s47}
F_{2}\approx \frac{1}{4}%
\left\vert f_{cg}f_{ce}+f_{rg}\left( f_{ce}\right) ^{2}\right\vert
^{2}.
\end{equation}%
Here, the parameters $f_{ce}$ and $f_{rg}$ are referred to
Eqs.~(\ref{eq:s39}) and (\ref{eq:s43}), however the parameter
$f_{cg}$ is given as
\begin{equation}\label{eq:s48}
f_{cg}=\left( 1-\frac{2\Omega ^{2}}{\delta ^{2}}\right) \sin \left[ \frac{%
\pi }{2}\left( 1-\frac{\Omega ^{2}}{2\delta ^{2}}\right) \right].
\end{equation}

\begin{figure}
\includegraphics[bb=10 5 385 285, width=4.2 cm, clip]{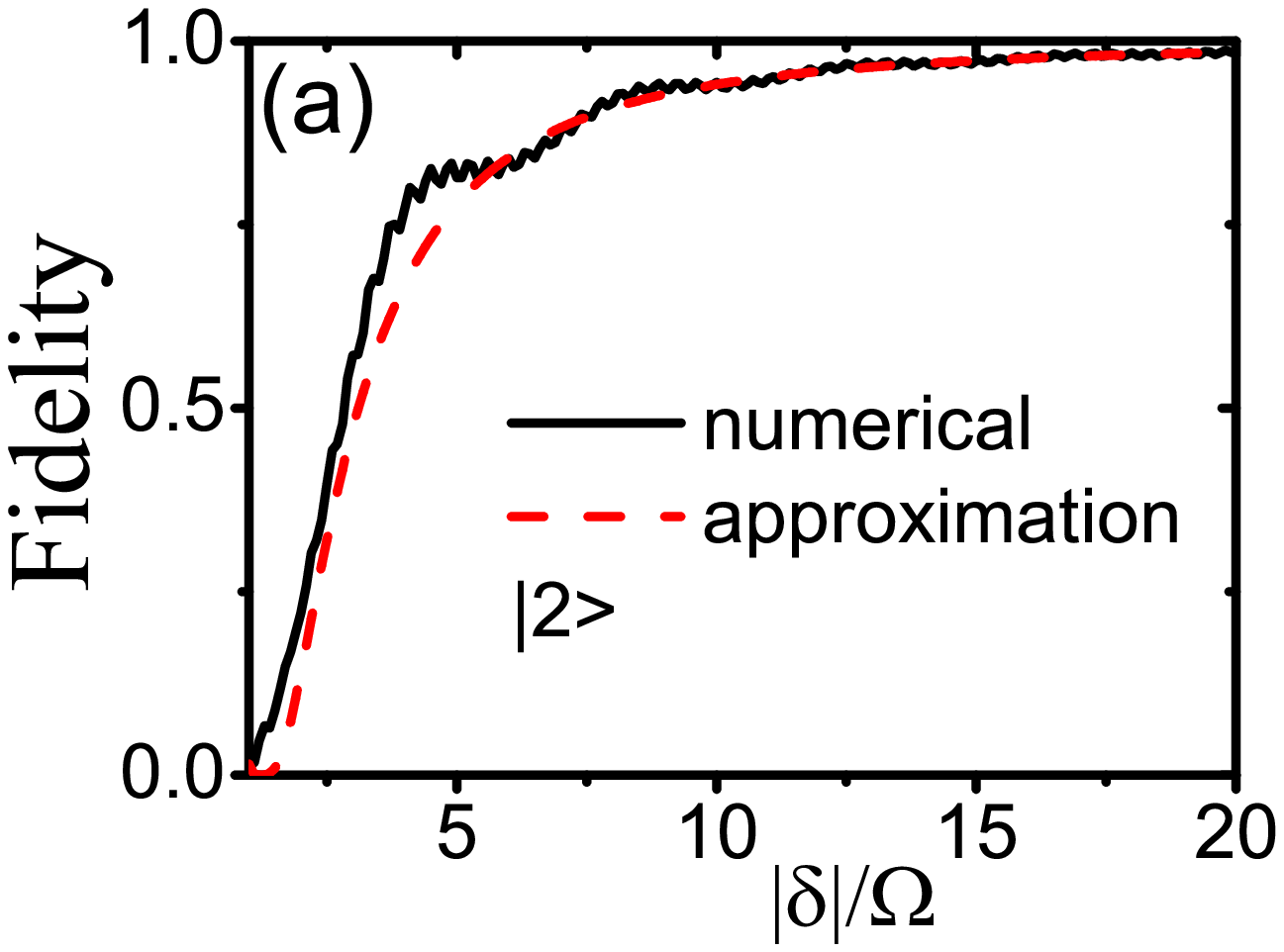}
\includegraphics[bb=10 5 385 285, width=4.2 cm, clip]{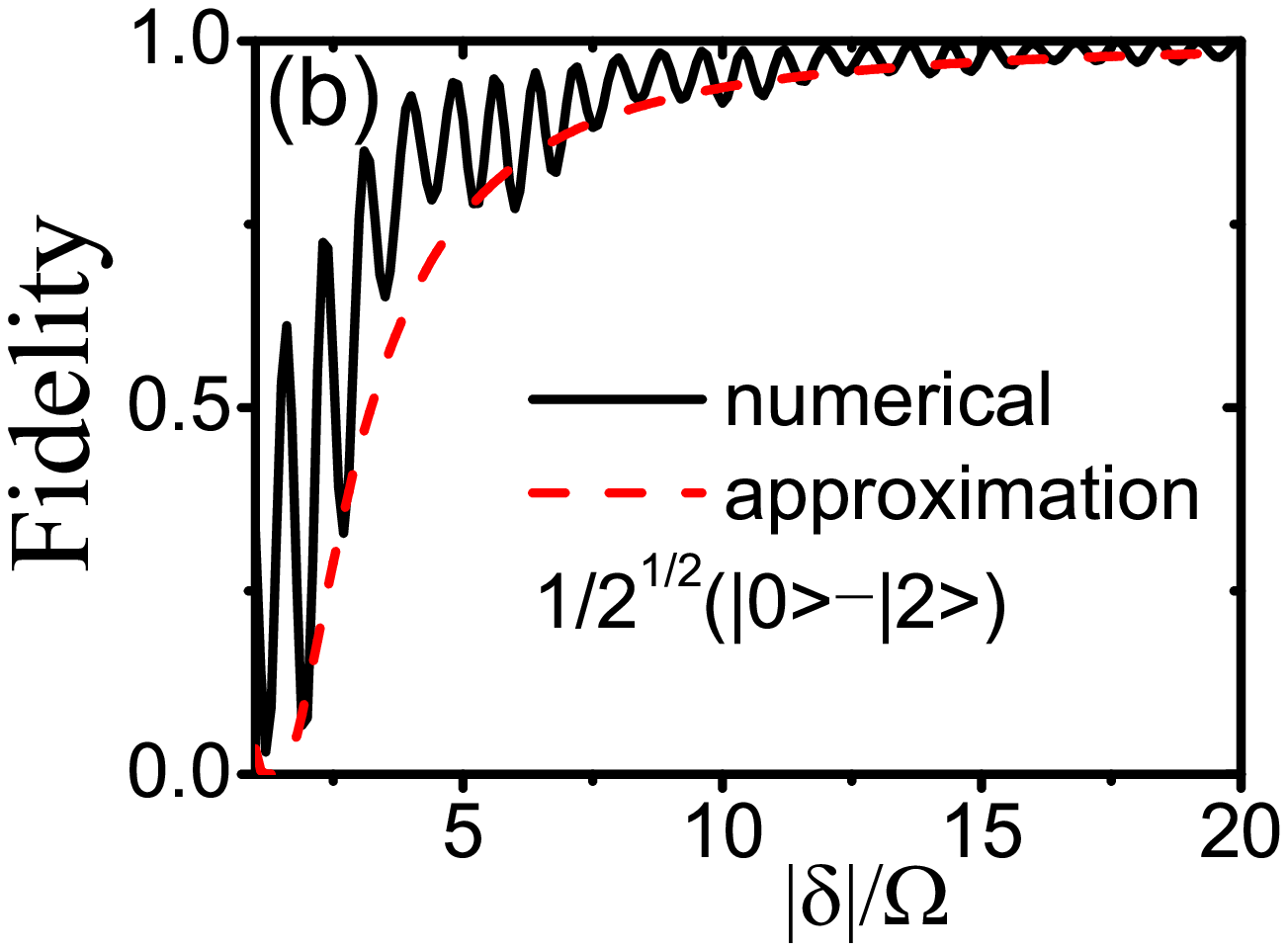}
%\centerline{\epsfxsize=8.5cm\epsfbox{fig1.eps}}
\caption[]{(Color online) Fidelities are plotted as a function of
$|\delta|/\Omega$ for preparing states $|2\rangle$ in (a) and
$(|0\rangle -|2\rangle)/\sqrt{2}$ in (b) using both numerical (black
solid curve) and approximately analytical (red dash curve) results
for $\eta=0.1$. }\label{fig3}
\end{figure}

In Figs.~\ref{fig3}(a) and (b), $F_{1}$
and $F_{2}$ are numerically calculated and also compared with the
approximated solutions in Eqs.~(\ref{eq:s46}) and (\ref{eq:s47}). We
find that the fidelities tend to one when $|\delta|/\Omega
> 20$, moreover, the fidelity is bigger than $0.9$ when
$|\delta|/\Omega>10$. Thus, it is clear that the large anharmonicity $\delta$
corresponds to good two-level approximation.

\begin{figure}
\includegraphics[bb=20 8 410 288, width=4.2 cm, clip]{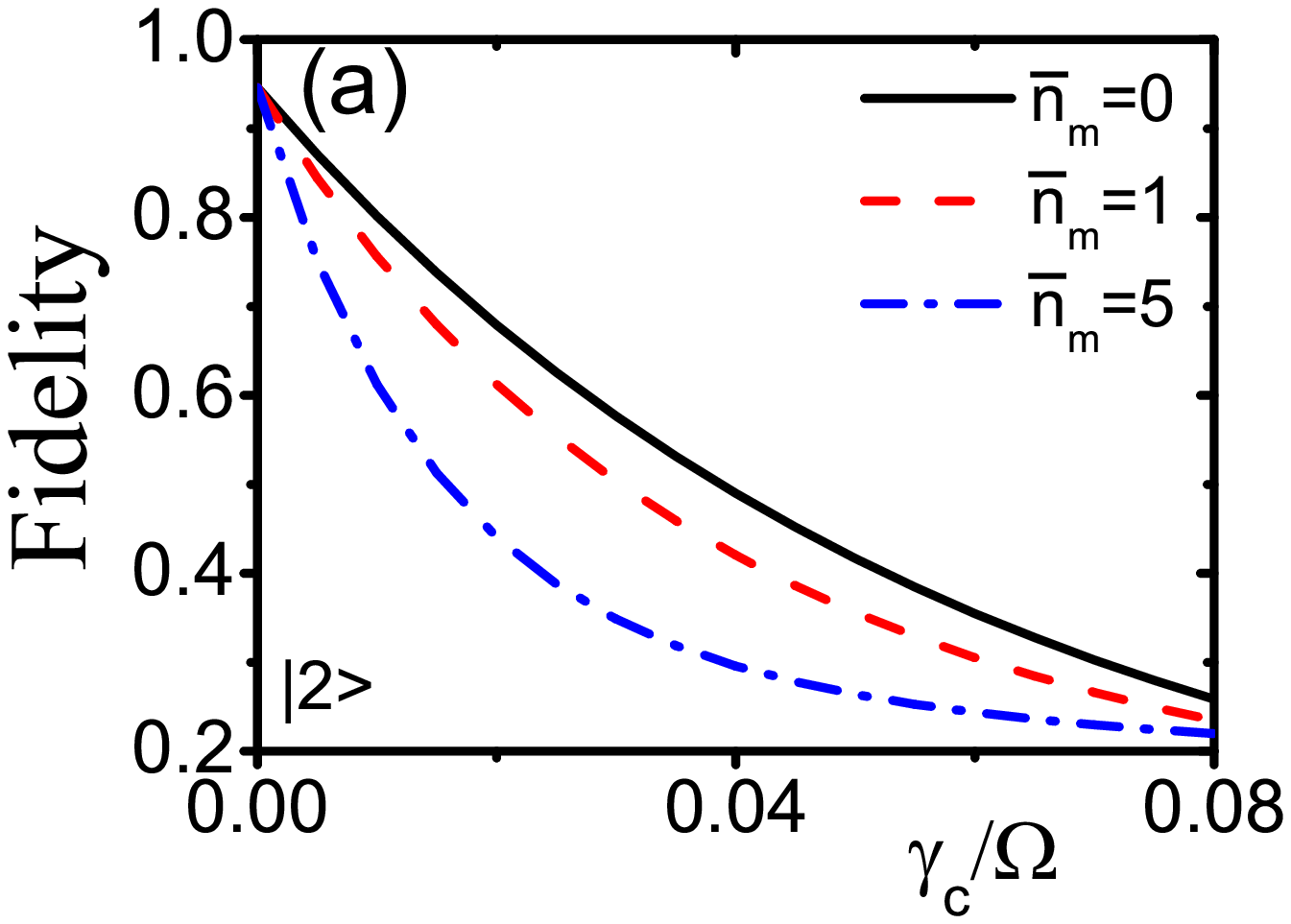}
\includegraphics[bb=20 8 410 288, width=4.2 cm, clip]{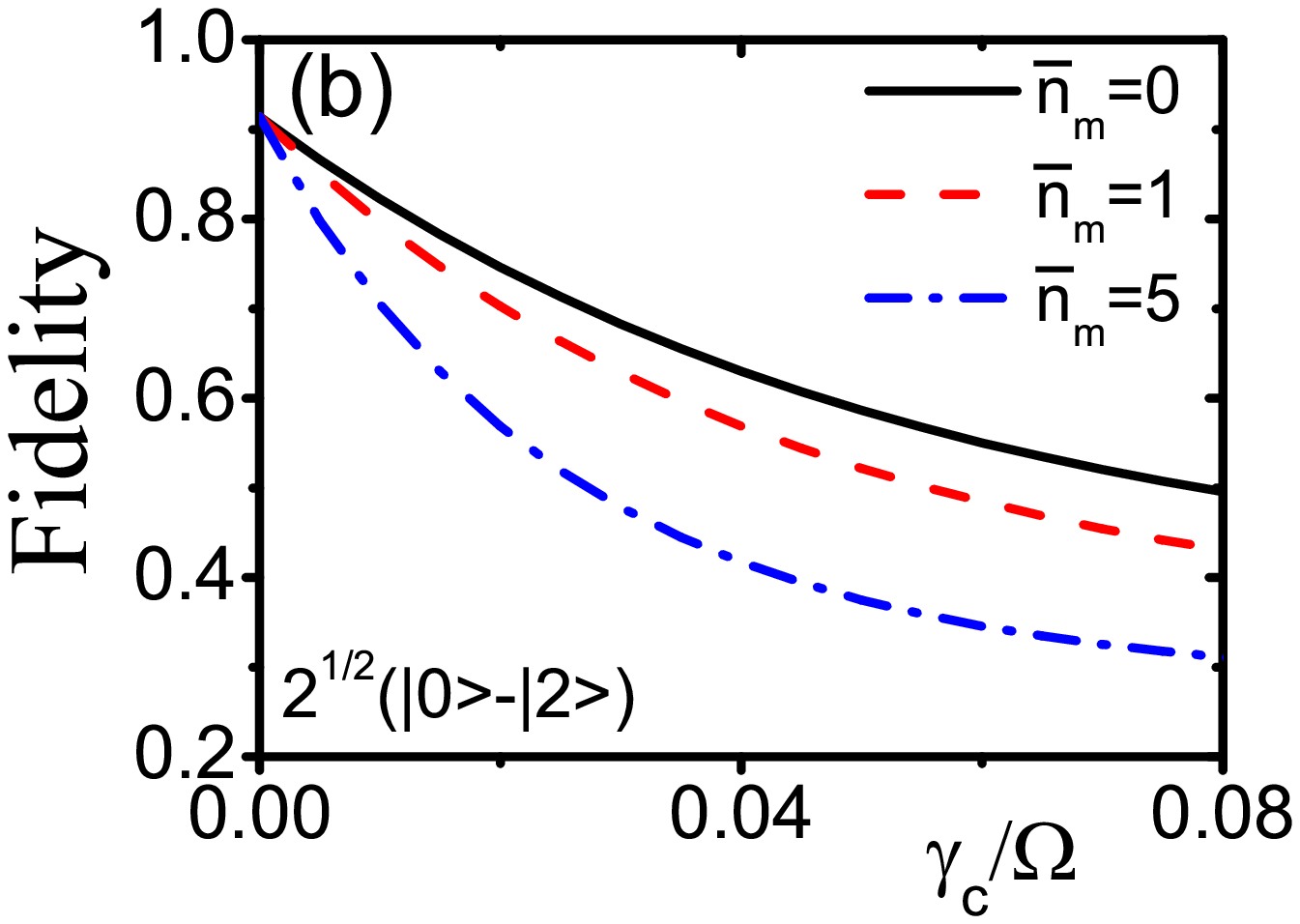}
%\centerline{\epsfxsize=8.5cm\epsfbox{fig1.eps}}
\caption[]{(Color online) Fidelities are plotted for preparing
states $|2\rangle$ in (a) and $(|0\rangle -|2\rangle)/\sqrt{2}$ in
(b) as a function of the cavity decay rate $\gamma _{c}/\Omega$ with different temperatures ($\bar{n}_{m}=0,1,5$) for
$\gamma _{m}=\gamma _{c}/10$, $|\delta|/\Omega=10$,
and $\eta=0.1$.}\label{fig4} \end{figure}

\section{Environmental effect on phonon states preparation}

We now study the environmental effect on the phonon state preparation. After the environmental effect is included, the
dynamical evolution of the reduced density operator $\rho(t)$ of the
optomechanical system can be described
by using the master equation~\cite{Carmichael}
\begin{eqnarray}\label{eq:8}
\frac{d \rho }{d t} &=&\frac{1}{i\hbar }\left[ H_{\rm eff},\rho \right] +
\frac{\gamma_{c}}{2}\left( 2\tilde{a}\rho \tilde{a}^{\dag }-\tilde{a}^{\dag }\tilde{a}\rho -\rho
\tilde{a}^{\dag }\tilde{a}\right)  \nonumber\\
&&+\frac{\gamma _{m}}{2}\left( 2\tilde{b}\rho \tilde{b}^{\dag }-\tilde{b}^{\dag }\tilde{b}\rho -\rho
\tilde{b}^{\dag }\tilde{b}\right)  \nonumber\\
&&+\gamma _{m} \bar{n}_{m}\left( \tilde{b}\rho \tilde{b}^{\dag }+\tilde{b}^{\dag }\rho
\tilde{b}-\tilde{b}^{\dag }\tilde{b}\rho -\rho \tilde{b} \tilde{b}^{\dag }\right),
\end{eqnarray}%
here, $\gamma _{m}$ is the decay rate of mechanical mode and $\bar{n}_{m}=
1/[\exp(\hbar\omega/k_{B} T)-1]$ is the thermal  phonon number of the mechanical resonator with the Boltzmann constant $k_{B}$ and the environmental temperature $T$. In Eq.~(\ref{eq:8}), we have set
\begin{equation}
\tilde{a}=UaU^{\dag }=ae^{-\eta\left( b^{\dag }-b\right) }
\end{equation}
and
\begin{equation}
\tilde{b}=UbU^{\dag }=b-\eta a^{\dag }a.
\end{equation}
When Eq.~(\ref{eq:8}) is written out, we have assumed that the single-photon energy of the cavity field is much bigger than the thermal excitation energy and the single-phonon energy of the mechanical resonator, i.e., $\hbar\omega_{c}\gg k_{B} T$ and $ \hbar\omega_{c}\gg \hbar\omega_{m}$, thus the thermal excitation on the cavity field is neglected under the condition of the low environmental temperature. In this case,  the environment of the cavity field is assumed at the zero temperature, but the environmental temperature of the mechanical resonator is assumed as a finite value $T$.

In the limit $\eta \ll 1$ and also for simplicity of the calculations, we can neglect the terms including, e.g. $\eta\gamma_{m}b \rho a^{\dagger}a$ and $O\left(\eta^{2}\right)$, thus Eq.~(\ref{eq:8}) is simplified to
\begin{eqnarray}\label{eq:8-2}
\frac{d \rho }{d t} & \approx &\frac{1}{i\hbar }\left[ \widetilde{H}_{\rm thr},\rho \right] +
\frac{\gamma_{c}}{2}\left( 2a\rho a^{\dag }-a^{\dag }a\rho -\rho
a^{\dag }a\right)  \nonumber\\
&&+\frac{\gamma _{m}}{2}\left( 2b\rho b^{\dag }-b^{\dag }b\rho -\rho
b^{\dag }b\right)  \nonumber\\
&&+\gamma _{m} \bar{n}_{m}\left( b\rho b^{\dag }+b^{\dag }\rho
b-b^{\dag }b\rho -\rho b b^{\dag }\right),
\end{eqnarray}%
with the photon operator given by
\begin{equation}
a\approx |g\rangle \langle e|+\sqrt{2}|e\rangle \langle e^{\prime }|
\end{equation}
in the basis of three lowest energy levels of the photon system.
The fidelity of the prepared phonon states can be calculated by~\cite{Uhlmann}
\begin{equation}\label{eq:9}
F=\left[ {\rm Tr}\left(\sqrt{\sqrt{\rho_{r}} \rho_{t}
\sqrt{\rho_{r}}}\right)\right]^{2},
\end{equation}
where $\rho_{t}$ is the density operator of the target phonon state
and $\rho_{r}$ is the reduced density operator of the mechanical
resonator obtained by numerically solving the master
equation~(\ref{eq:8-2}).

The fidelities, for
preparing states $|2\rangle$ and $(|0\rangle -|2\rangle)/\sqrt{2}$,
as a function of the cavity decay rate $\gamma _{c}/\Omega$ are shown in Fig.~\ref{fig4}. We find
that the fidelities decrease with the increase of the decay rates of
the cavity field and mechanical mode. To obtain the acceptable
fidelity of the prepared state, the decay rates should be much
smaller than the Rabi frequency of the external driven field, i.e.
$\gamma _{c} \ll \Omega$. Moreover, the fidelities of
the target states decrease with the increase of the thermal phonon
in the mechanical resonator.

\section{Discussions on experimental feasibility}
Let us now discuss the experimental feasibility of our proposal. (i) Similar to
the ground-state cooling of the optomechanical system, the
generation of arbitrary superpositions of phonon states relies on
the sideband excitations. This means that the frequency $\omega_{m}$
of the mechanical resonator and the decay rate $\gamma_{c}$ of the
cavity field have to satisfy the condition $\omega_{m}>\gamma_{c}$.
(ii) Our proposal should work at the single-photon strong coupling
regime as for the photon blockade ~\cite{blockade1,blockade2}  in
optomechanical systems, thus the coupling strength $g$ and the
frequency $\omega_{m}$ of the mechanical resonator should be larger
than the decay rates $\gamma_{c}$ and $\gamma_{m}$ of the mechanical
resonator and the cavity field, i.e., $\omega_{m},\, g \gg
\gamma_{c},\,\gamma_{m}$. Moreover, the nonlinear photon-photon
interaction strength $g^2/\omega_{m}$ should be bigger than the
decay rate $\gamma_{c}$ of the cavity field, i.e.,
$(g^2/\omega_{m})>\gamma_{c}$, such that the single-photon
excitation or photon blockade can be guaranteed and the two-level
approximation can be applied. (iii) Negligible information leakage
requires that the strength $\Omega$ of the classical driving field
should be smaller than the anharmoncity $ 2 g^2/\omega_{m}$ in the
carrier process. However, coherent transfer of excitations requires
that the excitation time $2\pi/\Omega$ of the cavity field should be much smaller than the
decay times $2\pi/\gamma_{c}$ of the cavity field for negligible temperature effect
and $2\pi/[(\bar{n}_{m}+1)\gamma_{m}]$ of the mechanical mode at the finite temperature. Here, the thermal phonon number
$\bar{n}_{m}$ is referred to Eq.~(\ref{eq:8}). (iv) As in trapped ion
systems~\cite{wei}, the big Lamb-Dicke parameter $\eta=g/\omega_{m}$
corresponds to the fast preparation of the multi-phonon states.
Therefore, the big $\eta$ is more desirable for our proposal.

\begin{widetext}
\begingroup
\squeezetable
\begin{table*}
\caption{\label{tab1} Summary of the Lamb-Dicke parameters
for current optomechanical experiments in microwave and optical domains. In the table,
$\omega_{c}$ and $\gamma_{c}$ are the frequency and the decay rate
of the cavity, the parameters $\omega_{m}$ and $\gamma_{m}$ are the frequency and
the decay rate of the mechanical resonator, $g$ is the
optomechanical coupling constant, and $\eta=g/\omega_{m}$ is the
Lamb-Dicke parameter. The values given in parentheses are the
expected experimental parameters achieved in the future for
realizing our proposal. $F_{1}$ and $F_{2}$ are the fidelities for
preparing states $|2\rangle$ and $(|0\rangle -|2\rangle)/\sqrt{2}$,
respectively. They are calculated at the zero temperature ($\bar{n}_{m}=0$) by using the parameters given in
the parentheses with the corresponding Rabi frequency $\Omega$. }
%\begin{ruledtabular}
\begin{tabular}{|l|l|l|l|l|l|l|l|l|l|l|l|l|c|c|}
\hline
System  & $\omega_{c}/2\pi~(\rm Hz)$     & \multicolumn{2}{|c|}{$\gamma_{c}/2\pi~(\rm Hz)$}
        & \multicolumn{2}{|c|}{$\omega_{m}/2\pi~(\rm Hz)$}      & \multicolumn{2}{|c|}{$\gamma_{m}/2\pi~(\rm Hz)$}
        & \multicolumn{2}{|c|}{$g/2\pi~(\rm Hz)$}               & \multicolumn{2}{|c|}{$\eta=g/\omega_{m} $}
        & $\Omega/2\pi~(\rm Hz)$         & $F_{1}$              & $F_{2}$   \\
\hline
Microwave cavity~ \cite{strong3}            & 7.47 G     & 170 K   & (1 K)   & 10.69 M   & (100 M)      & 30      & (10)    & 226    & (10 M)
& $2.11\times10^{-5}$      & (0.1)                  & 50 K     & 0.7359      & 0.8170 \\
Toroidal microcavity~\cite{Kippenberg}      & 385 T      & 7.1 M   & -       & 78 M      & -            & 10 K    & -       & 3.4 K  & -
& $4.36\times10^{-5}$      & -              &          &             & \\
Optomechanical crystals~\cite{Painter1}     & 195 T      & 500 M   & (0.1 M) & 3.68 G    & (10 G)       & 35 K    & (5 K)   & 910 K  & (1 G)
& $2.47\times10^{-4}$      & (0.1)                  & 5 M      & 0.7205      & 0.8108 \\
BEC~\cite{Esslinger}                        & 385 T      & 1.3 M   &
(0.1 M) & 15.1 K    & (10 M)       &         & (10)    & 0.39 M & (1
M)
& 25.828                   & (0.1)                  & 5 K      & 0.7017      & 0.8032 \\
Membrane~\cite{Harris}                      & 282 T      & 0.32 M  & -       & 134 K     & -            & 0.12    & -       & 2.68   & -
& $ 2\times10^{-5}$        & -              &          &             & \\
F-P cavity~\cite{Aspelmeyer}                & 282 T      & 215 K   & -       & 947 K     & -            & 140     & -       & 2.7    & -
& $2.85\times10^{-6}$      & -              &          &             & \\
Zipper cavity~\cite{Painter2}               & 194 T      & 6 G     & -       & 7.9 M     & -            & 98.75 K & -       & 599 K  & -
& $7.58\times10^{-2}$      & -              &          &             & \\
Double-wheel microcavity~\cite{Lipson}      & 190 T      & 10 G    & -       & 8.05 M    & -            & 2.01 M  & -       & 732 K  & -
& $9.09\times10^{-2}$      & -              &          &             & \\
\hline
\end{tabular}
%\end{ruledtabular}
\end{table*}
\endgroup
\end{widetext}

Based on above discussions, we estimate experimental parameters for
our goal. In Table~\ref{tab1}, we have summarized the parameters
used for current experiments of optomechanical systems. We find that
the promising candidates for realizing our proposal might be the
optomechanical crystal devices~\cite{Painter1}, the ultracold atoms
in optical resonators~\cite{strong1,strong2} and superconducting
circuits~\cite{strong3}. However, the parameters, e.g., the coupling strength $g$ and the frequency $\omega_m$ of the mechanical resonator, used for current experiments~~\cite{Painter1,strong1,strong2,strong3}
still need to be improved several orders of
magnitude for our proposal. The improvements for decay rates $\gamma_{c}$ and $\gamma_{m}$  might be achieved by further increasing the quality factors of the optical cavity and mechanical resonator. However, the coupling constant $g$ might be effectively increased by adding some impurities in the optomechanical systems~\cite{Ian-opto}. In the zero temperature with $\bar{n}_{m}=0$, we
calculated the optimal fidelities $F_{1}$ and $F_{2}$ for the target states $|2\rangle$ and $(|0\rangle
-|2\rangle)/\sqrt{2}$ using further possible parameters, as shown  in parentheses of
Table~\ref{tab1}, we find that the fidelities $F_{1}>0.7$ and $F_{2}>0.8$ can be achieved with these parameters.

\section{Conclusions}
In summary, we have proposed a method to synthesize
arbitrary non-classical single-mode phonon states in optomechanical systems by combining
photon blockade and sideband excitations. Similar to the photon
blockade ~\cite{blockade1,blockade2}, our proposal relies on the
single-photon strong coupling condition such that the two-level
approximation for the cavity field can be made in the optomechanical
systems. Our proposal opens up a possible way to deterministically engineer
arbitrary nonclassical single-mode phonon states on chip, and can also be generalized to engineering of multi-mode entangled phonon states in optomechanical systems~\cite{entanglephonon}.  The parameters, taken for calculations of the fidelities by us,
are ambitious, but we hope that our proposal can be realized in the near future with significant
improvement of the experiments.

\section{Acknowledgement}
Y.X.L. is supported by the National Natural Science Foundation of
China under Nos. 10975080 and 61025022. J.Z. is supported by the
NSFC under Nos. 61174084, 61134008.

\appendix

\section{The time evolution operators}\label{Evolution}

To show how the time evolution operators of the carrier, red and blue
sideband precesses can be derived, it is convenient to work in the interaction picture by using
\begin{equation}\label{A1}
V=e^{iH_{0}t/\hbar }H_{\text{th}}e^{-iH_{0}t/\hbar }
\end{equation}
with
\begin{equation*}
H_{0}=\hbar\omega_{m}b^{\dagger}b+(\hbar\omega\sigma_{z})/2,
\end{equation*}
here, the Hamiltonian $H_{\text{th}}$ is given in Eq.~(\ref{eq:3}). Equation~(\ref{A1}) can be further expressed as
\begin{equation}\label{eq:s5}
V=\hbar \Omega \sigma _{+}e^{\left( -\frac{1}{2}\eta ^{2}-i\phi _{d}\right) }%
\underset{j,j^{\prime }}{\sum }\frac{\left( -1\right) ^{j^{\prime
}}\eta
^{\left( j+j^{\prime }\right) }b^{\dag j}b^{j^{\prime }}}{j!j^{\prime }!}%
e^{-i \Delta t}+\text{h.c.},
\end{equation}
where $\Delta= \omega _{d}-\omega +\left( j^{\prime }-j\right) \omega_{m}$. Using the Schr\"{o}dinger  equation,
the wave function at any time $t$ can be given by
\begin{equation}\label{eq:s6}
\left\vert \psi \left( t\right) \right\rangle =U\left( t\right)
\left\vert \psi \left( 0\right) \right\rangle ,
\end{equation}%
where $U\left( t\right) =\exp \left( -iVt/\hbar \right)$ is the time evolution operator.
By using the identity operator
\begin{equation}\label{eq:s7}
\underset{n=0}{\overset{+\infty }{\sum }}\underset{i=g,e}{\sum
}\left\vert n,i\right\rangle \left\langle i,n\right\vert =1 ,
\end{equation}%
we can write out the time evolution operator $U\left( t\right)$ explicitly for different resonant conditions. If the
cavity is driven by a red-sideband excitation with the frequency of
the driving field $\omega _{d}=\omega -k\omega _{m}$, then the time
evolution operator is $U_{k}^{r}\left( t^{r}\right) =\sum^{+\infty }_{n=0} U_{n,k}^{r}\left( t^{r}\right)$ for the time
interval $t^{r}$, where $U_{n,k}^{r}\left( t^{r}\right)$ is given by
\begin{widetext}
\begin{equation}\label{eq:s8}
U_{n,k}^{r}\left( t^{r}\right) =\left\{
\begin{array}{ll}
\begin{array}{l}
\left\vert n,g\right\rangle \left\langle n,g\right\vert
+\left[ \cos \left( \Omega _{n,k}t^{r}\right) \left\vert
n,e\right\rangle -i\left( -1\right) ^{k}e^{i\phi _{r}}\sin \left(
\Omega _{n,k}t^{r}\right) \left\vert n+k,g\right\rangle \right]
\left\langle n,e\right\vert
\end{array}
& n<k, \\
\begin{array}{l}
\left[ \cos \left( \Omega _{n-k,k}t^{r}\right) \left\vert
n,g\right\rangle -i\left( -1\right) ^{k}e^{-i\phi _{r}}\sin \left(
\Omega _{n-k,k}t^{r}\right)
\left\vert n-k,e\right\rangle \right] \left\langle n,g\right\vert  \\
+\left[ \cos \left( \Omega _{n,k}t^{r}\right) \left\vert
n,e\right\rangle -i\left( -1\right) ^{k}e^{i\phi _{r}}\sin \left(
\Omega _{n,k}t^{r}\right) \left\vert n+k,g\right\rangle \right]
\left\langle n,e\right\vert
\end{array}
& n\geqslant k,%
\end{array}%
\right.
\end{equation}
with the Rabi frequency
\begin{equation}\label{eq:s9}
\Omega _{n,k}=\Omega \eta ^{k}e^{-\frac{1}{2}\eta
^{2}}\sqrt{\frac{\left( n+k\right)
!}{n!}}\overset{n}{\underset{j=0}{\sum }}\frac{\left( -1\right)
^{j}\eta ^{2j}}{j!\left( j+k\right) !}\frac{n!}{\left( n-j\right) !}.
\end{equation}
When the cavity is driven by blue-sideband excitation with the
frequency $\omega _{d}=\omega +k\omega _{m}$, then the time
evolution operator is $U_{k}^{b}\left( t^{b}\right) =\sum^{+\infty }_{n=0} U_{n,k}^{b}\left( t^{b}\right)$ with the time
interval $t^b$, where $U_{n,k}^{b}\left( t^{b}\right)$ is given as
\begin{equation}\label{eq:s10}
U_{n,k}^{b}\left( t^{b}\right) =\left\{
\begin{array}{ll}
\begin{array}{l}
\left[ \cos \left( \Omega _{n,k}t^{b}\right) \left\vert n,g\right\rangle
-ie^{-i\phi _{b}}\sin \left( \Omega _{n,k}t^{b}\right) \left\vert
n+k,e\right\rangle \right] \left\langle n,g\right\vert
+\left\vert n,e\right\rangle \left\langle n,e\right\vert
\end{array}
& n<k, \\
\begin{array}{l}
\left[ \cos \left( \Omega _{n,k}t^{b}\right) \left\vert n,g\right\rangle
-ie^{-i\phi _{b}}\sin \left( \Omega _{n,k}t^{b}\right) \left\vert
n+k,e\right\rangle \right] \left\langle n,g\right\vert \\
+\left[ \cos \left( \Omega _{n-k,k}t^{b}\right) \left\vert
n,e\right\rangle -ie^{i\phi _{b}}\sin \left( \Omega _{n-k,k}t^{b}\right)
\left\vert n-k,g\right\rangle \right] \left\langle n,e\right\vert
\end{array}
& n\geqslant k.%
\end{array}%
\right.
\end{equation}
Finally, if the driving filed is resonant with the two-lowest energy
levels of the cavity field, e.g. $\omega _{d}=\omega$, then the
carrier process occurs and the time evolution operator is $U^{c}\left( t^{c}\right) =\sum^{+\infty }_{n=0} U_{n,0}^{c}\left( t^{c}\right)$ with the time interval $t^c$,
where $U_{n,0}^{c}\left( t^{c}\right)$ is given by
\begin{eqnarray}\label{eq:s11}
U_{n,0}^{c}\left( t^{c}\right) &=& \left[ \cos \left( \Omega
_{n,0}t^{c}\right) \left\vert n,g\right\rangle -ie^{-i\phi _{c}}\sin
\left( \Omega _{n,0}t^{c}\right) \left\vert
n,e\right\rangle \right] \left\langle n,g\right\vert  \nonumber\\
&&+\left[ \cos \left( \Omega _{n,0}t^{c}\right) \left\vert
n,e\right\rangle -ie^{i\phi _{c}}\sin \left( \Omega _{n,0}t^{c}\right)
\left\vert n,g\right\rangle \right] \left\langle n,e\right\vert.
\end{eqnarray}
\end{widetext}
Under different resonant conditions, dynamical
evolutions of the system are governed by the red sideband
excitation, blue sideband excitation, and carrier process with the time
evolution operators given in Eqs.~(\ref{eq:s8})-(\ref{eq:s11}),
respectively.

\section{Leakage effect}\label{Leakage}

In order to analyze the effect of the third level
$|e^{\prime}\rangle$ of the cavity field
on the fidelity for preparing the phonon states $|2\rangle $ and $
(|0\rangle -|2\rangle )/\sqrt{2}$ from the initial state
$|0,g\rangle$, let us assume that the wavefunction of the optomechanical system  for the carrier and sideband excitations can be written as
\begin{eqnarray}\label{eq:s22}
|\varphi (t)\rangle  &=&c_{0,g}(t)|0,g\rangle +c_{1,g}(t)|1,g\rangle
+c_{2,g}(t)|2,g\rangle   \nonumber \\
&&+c_{0,e}(t)|0,e\rangle +c_{1,e}(t)|1,e\rangle
+c_{2,e}(t)|2,e\rangle
\notag \\
&&+c_{0,e^{\prime}}(t)|0,e^{\prime }\rangle +c_{1,e^{\prime
}}(t)|1,e^{\prime }\rangle +c_{2,e^{\prime }}(t)|2,e^{\prime
}\rangle, \nonumber\\
\end{eqnarray}
at the time $t$, where sideband excitation includes only the single-phonon process.

For the carrier process, $\omega
_{d}=\omega$ and there is no phonon exchange when the photon is excited, thus there is no transition between phonon Fock states with different phonon numbers and the coefficients $c_{m,i}(t)$ in Eq.~(\ref{eq:s22}) satisfy the dynamical
equations
\begin{widetext}
\begin{equation} \label{eq:s24}
i\partial _{t}\left(
\begin{array}{c}
\widetilde{c}_{0,g} \\
\widetilde{c}_{0,e} \\
\widetilde{c}_{0,e^{\prime }} \\
\widetilde{c}_{1,g} \\
\widetilde{c}_{1,e} \\
\widetilde{c}_{1,e^{\prime }} \\
\widetilde{c}_{2,g} \\
\widetilde{c}_{2,e} \\
\widetilde{c}_{2,e^{\prime }}%
\end{array}%
\right) =\left(
\begin{array}{ccc|ccc|ccc}
0 & \Omega  & 0 & 0 & 0 & 0 & 0 & 0 & 0 \\
\Omega  & 0 & \sqrt{2}\Omega  & 0 & 0 & 0 & 0 & 0 & 0 \\
0 & \sqrt{2}\Omega  & \delta  & 0 & 0 & 0 & 0 & 0 & 0 \\
\hline
0 & 0 & 0 & 0 & \Omega  & 0 & 0 & 0 & 0 \\
0 & 0 & 0 & \Omega  & 0 & \sqrt{2}\Omega  & 0 & 0 & 0 \\
0 & 0 & 0 & 0 & \sqrt{2}\Omega  & \delta  & 0 & 0 & 0 \\
\hline
0 & 0 & 0 & 0 & 0 & 0 & 0 & \Omega  & 0 \\
0 & 0 & 0 & 0 & 0 & 0 & \Omega  & 0 & \sqrt{2}\Omega  \\
0 & 0 & 0 & 0 & 0 & 0 & 0 & \sqrt{2}\Omega  & \delta
\end{array}%
\right) \left(
\begin{array}{c}
\widetilde{c}_{0,g} \\
\widetilde{c}_{0,e} \\
\widetilde{c}_{0,e^{\prime }} \\
\widetilde{c}_{1,g} \\
\widetilde{c}_{1,e} \\
\widetilde{c}_{1,e^{\prime }} \\
\widetilde{c}_{2,g} \\
\widetilde{c}_{2,e} \\
\widetilde{c}_{2,e^{\prime }}%
\end{array}%
\right),
\end{equation}%
where  $m=0,\,1,\,2$ denote the phonon states and
$i=g,\,e,\, e^{\prime}$ denote the photon states. In Eq.~(\ref{eq:s24}), we have also used the relations
\begin{eqnarray}
\widetilde{c}_{m,g}&=&c_{m,g}\exp \left( im\omega _{m}t\right), \nonumber \\
\widetilde{c}_{m,e}&=&c_{m,e}\exp \left[ i\left( \omega _{d}+m\omega_{m}\right) t\right],\\
\widetilde{c}_{m,e^{\prime }}&=&c_{m,e^{\prime }}\exp \left[ i\left( 2\omega _{d}+m\omega _{m}\right) t\right]. \nonumber
\end{eqnarray}
For the red sideband excitation with
 $\omega _{d}=\omega -\omega _{m}$, we can also have
\begin{equation}\label{eq:s26}
i\partial _{t}\left(
\begin{array}{c}
\widetilde{c}_{0,g} \\
\widetilde{c}_{1,g} \\
\widetilde{c}_{0,e} \\
\widetilde{c}_{2,g} \\
\widetilde{c}_{1,e} \\
\widetilde{c}_{0,e^{\prime }} \\
\widetilde{c}_{2,e} \\
\widetilde{c}_{1,e^{\prime }} \\
\widetilde{c}_{2,e^{\prime }}%
\end{array}%
\right) =\left(
\begin{array}{c|cc|ccc|cc|c}
0 & 0 & 0 & 0 & 0 & 0 & 0 & 0 & 0 \\
\hline
0 & 0 & -\eta \Omega  & 0 & 0 & 0 & 0 & 0 & 0 \\
0 & -\eta \Omega  & 0 & 0 & 0 & 0 & 0 & 0 & 0 \\
\hline
0 & 0 & 0 & 0 & -\eta \Omega \sqrt{2} & 0 & 0 & 0 & 0 \\
0 & 0 & 0 & -\eta \Omega \sqrt{2} & 0 & -\eta \Omega \sqrt{2} & 0 & 0 & 0 \\
0 & 0 & 0 & 0 & -\eta \Omega \sqrt{2} & \delta  & 0 & 0 & 0 \\
\hline
0 & 0 & 0 & 0 & 0 & 0 & 0 & -2\eta \Omega  & 0 \\
0 & 0 & 0 & 0 & 0 & 0 & -2\eta \Omega  & \delta  & 0 \\
\hline
0 & 0 & 0 & 0 & 0 & 0 & 0 & 0 & \delta
\end{array}%
\right) \left(
\begin{array}{c}
\widetilde{c}_{0,g} \\
\widetilde{c}_{1,g} \\
\widetilde{c}_{0,e} \\
\widetilde{c}_{2,g} \\
\widetilde{c}_{1,e} \\
\widetilde{c}_{0,e^{\prime }} \\
\widetilde{c}_{2,e} \\
\widetilde{c}_{1,e^{\prime }} \\
\widetilde{c}_{2,e^{\prime }}%
\end{array}%
\right).
\end{equation}%
\end{widetext}
The dynamical evolutions of the system in different conditions can
be obtained by numerically solving Eqs.~(\ref{eq:s24}) and
(\ref{eq:s26}). However, we can approximately give an analytical solution
by using the method in Ref.~\cite{AminLTP} as shown below.

The left matrixes of the right hand of the dynamical equations in
Eqs.~(\ref{eq:s24}) and (\ref{eq:s26}) have block diagonal form,
thus the calculations on total solutions of Eqs.~(\ref{eq:s24}) and
(\ref{eq:s26}) can be greatly reduced. It is clear that the problem
of solving Eq.~(\ref{eq:s24}) is equivalent to that of solving three
linear differential equations
\begin{eqnarray}\label{eq:s27}
i\partial _{t}\widetilde{c}_{G} &=& A \widetilde{c}_{E}, \nonumber \\
i\partial _{t}\widetilde{c}_{E} &=& A \widetilde{c}_{G}+ B \widetilde{c}_{E^{\prime }}, \\
i\partial _{t}\widetilde{c}_{E^{\prime }} &=& B \widetilde{c}_{E}+\delta \widetilde{c}_{E^{\prime }}, \nonumber
\end{eqnarray}%
where $A=\Omega$, $B=\sqrt{2} \Omega$, the subscript $G (=m,g)$
denotes that the cavity field is in the ground state with $m$
phonons. The subscript $E (=m,e)$ or $E^{\prime} (=m,e^{\prime})$
denote that the cavity field is in the first or the second excited
state with $m$ phonons. Here, $m$ takes values $0,\,1$ and $2$, which correspond to three different
block matrixes.

In the condition $\delta \gg \left\vert A \right\vert , \left\vert B
\right\vert $, the general solutions of Eq.~(\ref{eq:s27}) can be
given as
\begin{eqnarray}
\widetilde{c}_{G}(t) &=&\sum ^{3}_{n=1} c_{n}\exp \left( -i\omega_{n}t\right),\nonumber \\
\widetilde{c}_{E}(t) &=&\frac{1}{A}\sum ^{3}_{n=1} \omega_{n}c_{n}\exp \left( -i\omega _{n}t\right), \label{eq:B6}\\
\widetilde{c}_{E^{\prime }}(t) &=&\frac{1}{ A B }\sum ^{3}_{n=1}\left( \omega _{n}^{2}-A^{2}\right) c_{n}\exp \left( -i\omega _{n}t\right),\nonumber
\end{eqnarray}
with
\begin{eqnarray}
\omega _{1}&=&\frac{\Omega _{R}}{2}-\frac{ B^{2}}{2\delta }, \nonumber\\
\omega _{2}&=&-\frac{\Omega _{R}}{2}-\frac{ B^{2}}{2\delta },\\
\omega _{3}&=&\delta \left( 1+\frac{ B^{2}}{\delta ^{2}}\right), \nonumber
\end{eqnarray}
with the parameter $\Omega _{R}$, defined as
\begin{equation}
\Omega _{R}=2\left\vert A \right\vert \left( 1-\frac{ B^{2} }{2\delta ^{2}}+\frac{B^{4}}{8 A^{2}\delta ^{2}}\right).
\end{equation}
The coefficients $c_{n}$ in Eq.~(\ref{eq:B6}) with $n=1,2,3$ can be determined by the
initial condition.

If the cavity field and the mechanical mode are initially in the
ground state $|g\rangle$ and $m$-phonon state $|m\rangle$,
respectively, i.e., the initial state of the system is
$|m,g\rangle$, then we have $\widetilde{c}_{G}(0)=1$,
$\widetilde{c}_{E}(0)=\widetilde{c}_{E^{\prime }}(0)=0$, and thus
the coefficients $c_{n}$ satisfy following linear
equations
\begin{eqnarray}
c_{1}+c_{2}+c_{3} &=&1 , \nonumber\\
\omega _{1}c_{1}+\omega _{2}c_{2}+\omega _{3}c_{3} &=&0 , \\
\omega _{1}^{2}c_{1}+\omega _{2}^{2}c_{2}+\omega _{3}^{2}c_{3} &=& A^{2}. \nonumber
\end{eqnarray}%
In the condition $\delta \gg \left\vert A \right\vert,\left\vert B \right\vert $, we obtain
\begin{eqnarray}
c_{1} &\approx& \frac{1}{2}+\frac{ B^{2}}{4\left\vert A \right\vert \delta }, \nonumber\\
c_{2} &\approx& \frac{1}{2}-\frac{ B^{2}}{4\left\vert A \right\vert \delta }, \nonumber\\
c_{3} &\approx& 0 .
\end{eqnarray}
Thus, we have the solutions
\begin{eqnarray}\label{eq:s34}
\widetilde{c}_{G}(t) &\approx &\cos \left( \frac{\Omega_{R}}{2}t\right) -i\frac{B^{2}}{2\left\vert A \right\vert \delta }%
\sin \left( \frac{\Omega _{R}}{2}t\right), \nonumber\\
\widetilde{c}_{E}(t) &\approx &-i\frac{\left\vert A \right\vert }{A}\left( 1-\frac{ B^{2}}{2\delta^{2}}-\frac{ B^{4}}{8 A^{2}\delta ^{2}}\right) \sin \left( \frac{\Omega_{R}}{2}t\right), \nonumber\\
\widetilde{c}_{E^{\prime }}(t) &\approx & i \frac{\left\vert A
\right\vert}{A}\frac{B}{\delta }\sin \left( \frac{\Omega _{R}}{2}t\right),
\end{eqnarray}
when the whole system is initially in the ground state. Similarly, if the system is initially in the state $|m,e\rangle$, i.e.,
$\widetilde{c}_{E}(0)=1$, $\widetilde{c}_{G}(0)=\widetilde{c}_{E^{\prime }}(0)=0$,
then we have
\begin{eqnarray}
c_{1}+c_{2}+c_{3} &=& 0 ,\nonumber\\
\omega _{1}c_{1}+\omega _{2}c_{2}+\omega _{3}c_{3} &=& A ,\\
\omega _{1}^{2}c_{1}+\omega _{2}^{2}c_{2}+\omega _{3}^{2}c_{3} &=& 0 .\nonumber
\end{eqnarray}
By solving above linear equations in the condition $\delta \gg \left\vert A \right\vert,\left\vert B \right\vert $, we can have
\begin{eqnarray}
c_{1} &\approx &\frac{A}{2\left\vert A \right\vert }\left( 1-\frac{ B^{2}}{\delta^{2}}\right) ,\nonumber\\
c_{2} &\approx &-\frac{A}{2\left\vert A \right\vert }\left( 1-\frac{ B^{2}}{\delta^{2}}\right) ,\\
c_{3} &\approx &\frac{ A B^{2}}{\delta ^{3}} .\nonumber
\end{eqnarray}
Then, we have the solutions
\begin{eqnarray}\label{eq:s37}
\widetilde{c}_{G}(t) &\approx &-i\frac{A}{\left\vert A \right\vert }\left( 1-\frac{ B^{2}}{\delta
^{2}}\right) \sin \left( \frac{\Omega _{R}}{2}t\right), \nonumber\\
\widetilde{c}_{E}(t) &\approx &\left( 1-\frac{3 B^{2}}{2\delta^{2}}+\frac{ B^{4}}{8 A^{2}\delta^{2}}\right) \cos \left( \frac{\Omega _{R}}{2} t \right) , \nonumber\\
\widetilde{c}_{E^{\prime }}(t) &\approx &\frac{B}{\delta }\left\{ \cos \left[\left( \delta +\frac{3 B^{2}}{2\delta }\right) t \right] -\cos \left( \frac{\Omega _{R}}{2}t\right) \right\}\nonumber\\
&&-i\frac{B}{\delta }\sin \left[ \left( \delta +\frac{3 B^{2}}{2\delta }\right) t \right],
\end{eqnarray}
when the whole system is initially in the state $|m,e\rangle$.

Similarly, the solutions of Eq.~(\ref{eq:s26}) can be given by
solving the linear differential equations. The solutions of
the coefficients $\widetilde{c}_{2,g}$, $\widetilde{c}_{1,e}$ and
$\widetilde{c}_{0,e^{\prime }}$ can be given by solving
Eq.~(\ref{eq:s27}) with $A=B=-\eta\Omega\sqrt{2}$ and the subscripts are
taken as $G=(2,g)$, $E=(1,e)$, and $E^{\prime}=(0,e^{\prime})$. The
coefficient $\widetilde{c}_{0,g}$ ($\widetilde{c}_{2,e^{\prime }}$)
only depends on itself and initial condition. The coefficients $\widetilde{c}_{1,g}$ and
$\widetilde{c}_{0,e}$ ($\widetilde{c}_{2,e}$ and
$\widetilde{c}_{1,e^{\prime }}$) satisfy the following linear
differential equations
\begin{eqnarray}\label{eq:s28}
i\partial _{t}\widetilde{c}_{S} &=& A' \widetilde{c}_{X}, \nonumber \\
i\partial _{t}\widetilde{c}_{X} &=& A' \widetilde{c}_{S}+ B'
\widetilde{c}_{X},
\end{eqnarray}
for $A'=-\eta \Omega$ and $B'=0$ with the subscripts $S=(1,g)$
and $X=(0,e)$ [$A'=-2\eta \Omega$, $B'=\delta$ with the subscripts
$S=(2,e)$ and $X=(1,e^{\prime})$]. The general solutions of
Eq.~(\ref{eq:s28}) are given by
\begin{eqnarray}\label{eq:s29}
\widetilde{c}_{S}\left( t\right)  &=&\left\{ \widetilde{c}_{S}\left(
0\right) \left[ \cos \left( \frac{\Omega _{R^{\prime }}}{2}t\right) +\frac{%
iB^{\prime }}{\Omega _{R^{\prime }}}\sin \left( \frac{\Omega _{R^{\prime }}}{%
2}t\right) \right] \right.  \nonumber\\
&&\left. -i\frac{2A^{\prime }}{\Omega _{R^{\prime
}}}\widetilde{c}_{X}\left( 0\right) \sin \left( \frac{\Omega
_{R^{\prime }}}{2}t\right) \right\}
e^{-iB^{\prime }t/2} \nonumber\\
\widetilde{c}_{X}\left( t\right)  &=&\left\{ \widetilde{c}_{X}\left(
0\right) \left[ \cos \left( \frac{\Omega _{R^{\prime }}}{2}t\right) -\frac{%
iB^{\prime }}{\Omega _{R^{\prime }}}\sin \left( \frac{\Omega _{R^{\prime }}}{%
2}t\right) \right] \right.  \nonumber\\
&&\left. -i\frac{2A^{\prime }}{\Omega _{R^{\prime
}}}\widetilde{c}_{S}\left( 0\right) \sin \left( \frac{\Omega
_{R^{\prime }}}{2}t\right) \right\} e^{-iB^{\prime }t/2}
\end{eqnarray}%
where $\Omega _{R^{\prime }}=\sqrt{4A^{\prime 2}+B^{\prime 2}}$.


\begin{thebibliography}{99}

\bibitem{physrep}M. Poot and H. S. J. van der Zant, Phys. Rep.~\textbf{511}, 273 (2012).

\bibitem{review1}T. J. Kippenberg and K. J. Vahala, Science~\textbf{321}, 1172 (2008).

\bibitem{review2}M. Aspelmeyer, S. Gr\"oblacher, K. Hammerer, and N. Kiesel, J. Opt. Soc. Am. B~\textbf{27}, A189 (2010).

\bibitem{liu2004} Y. X. Liu, L. F. Wei, and F. Nori,  Europhys. Lett.~\textbf{67}, 941 (2004).

\bibitem{martinis-nature1}M. Hofheinz, E. M. Weig, M. Ansmann, R. C. Bialczak, E. Lucero, M. Neeley, A. D. O'Connell, H. Wang, J. M. Martinis, and A. N. Cleland, Nature~\textbf{454}, 310 (2008).

\bibitem{martinis-nature2}M. Hofheinz, H. Wang, M. Ansmann, R. C. Bialczak, E. Lucero, M. Neeley, A. D. O'Connell, D. Sank, J. Wenner, J. M. Martinis, and A. N. Cleland, Nature~\textbf{459}, 546 (2009).

\bibitem{Gardiner97}S. A. Gardiner, J. I. Cirac, and P. Zoller, Phys. Rev. A~\textbf{55}, 1683 (1997).

\bibitem{zheng}S. B. Zheng, Phys. Rev. A~\textbf{63}, 015801 (2001).

\bibitem{wei}L. F. Wei, Y. X. Liu, and F. Nori, Phys. Rev. A~\textbf{70}, 063801 (2004).

\bibitem{RMP}D. Leibfried, R. Blatt, C. Monroe, and D. Wineland, Rev. Mod. Phys.~\textbf{75}, 281 (2003).

\bibitem{ions1}D. M. Meekhof, C. Monroe, B. E. King, W. M. Itano, and D. J. Wineland, Phys. Rev. Lett.~\textbf{76}, 1796 (1996).

\bibitem{ions2}D. Leibfried, D. M. Meekhof, B. E. King, C. Monroe, W. M. Itano, and D. J. Wineland, Phys. Rev. Lett.~\textbf{77}, 4281 (1996).


\bibitem{armour}A. D. Armour, M. P. Blencowe, and K. C. Schwab, Phys. Rev. Lett.~\textbf{88}, 148301 (2002).


\bibitem{cleland and martinis}A. D. O'Connell, M. Hofheinz, M. Ansmann, R. C. Bialczak, M. Lenander, E. Lucero, M. Neeley, D. Sank, H. Wang, M. Weides, J. Wenner, J. M. Martinis, and A. N. Cleland, Nature~\textbf{464}, 697 (2010).

\bibitem{strong1}S. Gupta, K. L. Moore, K. W. Murch, and D. M. Stamper-Kurn, Phys. Rev. Lett.~\textbf{99}, 213601 (2007).

\bibitem{strong2}M. Eichenfield, J. Chan, R. M. Camacho, K. J. Vahala, and O. Painter, Nature~\textbf{462}, 78 (2009).

\bibitem{strong3}J. D. Teufel, D. Li, M. S. Allman, K. Cicak, A. J. Sirois, J. D. Whittaker, and R. W. Simmonds, Nature~\textbf{471}, 204 (2011).

\bibitem{strong4}J. C. Sankey, C. Yang, B. M. Zwickl, A. M. Jayich, and  J. G. E. Harris, Nature Phys.~\textbf{6}, 707 (2010).
%; N.E. Flowers-Jacobs,
%S. W. Hoch, J.C. Sankey, A. Kashkanova, A. M. Jayich, C. Deutsch, J.
%Reichel, and J.G.E. Harris, arXiv:1206.3558.

\bibitem{Imamoglu}A. Imamoglu, H. Schmidt, G. Woods, and M. Deutsch, Phys. Rev. Lett.~\textbf{79}, 1467 (1997).


\bibitem{blockade1}P. Rabl, Phys. Rev. Lett.~\textbf{107}, 063601 (2011).


\bibitem{blockade2}A. Nunnenkamp, K. Borkje, and S. M. Girvin, Phys. Rev. Lett.~\textbf{107}, 063602 (2011).

\bibitem{binghe}B. He, Phys. Rev. A~\textbf{85}, 063820 (2012).

\bibitem{JQLiao1}J. Q. Liao, H. K. Cheung, and C. K. Law, Phys. Rev. A~\textbf{85}, 025803 (2012).

\bibitem{JQLiao2}J. Q. Liao and F. Nori, Phys. Rev. A~\textbf{88}, 023853 (2013).


\bibitem{theory1} I. Wilson-Rae, N. Nooshi, W. Zwerger, and T. J. Kippenberg, Phys. Rev. Lett.~\textbf{99}, 093901 (2007).

\bibitem{theory2}F. Marquardt, J. P. Chen, A. A. Clerk, and S. M. Girvin, Phys. Rev. Lett.~\textbf{99}, 093902 (2007).

\bibitem{sideband1}Y. S. Park and H. L. Wang, Nature Phys.~\textbf{5}, 489 (2009).

\bibitem{sideband2}A. Schliesser, R. Riviere, G. Anetsberger, O. Arcizet, and T. J. Kippenberg, Nature Phys.~\textbf{4}, 415 (2008).

\bibitem{sideband3}A. Schliesser, O. Arcizet, R. Riviere, G. Anetsberger, and T. J. Kippenberg, Nature Phys.~\textbf{5}, 509 (2009).

\bibitem{sideband4}J. D. Teufel, T. Donner, D. Li, J. W. Harlow, M. S. Allman, K. Cicak, A. J. Sirois, J. D. Whittaker, K. W. Lehnert, and R. W. Simmonds, Nature~\textbf{475},  359 (2011).

\bibitem{Painter1}J. Chan, T. P. M. Alegre, A. H. Safavi-Naeini, J. T. Hill, A. Krause, S. Groblacher, M. Aspelmeyer, and O. Painter, Nature~\textbf{478}, 89 (2011).

\bibitem{Painter-new} A. H. Safavi-Naeini, J. Chan, J. T. Hill, S. Gr\"oblacher, H. Miao, Y. Chen, M. Aspelmeyer and O. Painter, New J. Phys.~\textbf{15}, 035007 (2013).


\bibitem{non1}W. Marshall, C. Simon, R. Penrose, and D. Bouwmeester, Phys. Rev. Lett.~\textbf{91}, 130401 (2003).

\bibitem{Pepper}B. Pepper, R. Ghobadi, E. Jeffrey, C. Simon, and D. Bouwmeester, Phys. Rev. Lett.~\textbf{109}, 023601 (2012).

\bibitem{kim}M. R. Vanner, M. Aspelmeyer, M. S. Kim, Phys. Rev. Lett.~\textbf{110}, 010504 (2013).



\bibitem{non2}F. Khalili, S. Danilishin, H. Miao, H. M\"uller-Ebhardt, H. Yang, and Y. Chen, Phys. Rev. Lett.~\textbf{105}, 070403 (2010).

\bibitem{non3}O. Romero-Isart, A. C. Pflanzer, F. Blaser, R. Kaltenbaek, N. Kiesel, M. Aspelmeyer, and J. I. Cirac, Phys. Rev. Lett.~\textbf{107}, 020405 (2011).

\bibitem{BlockleyEPL} C. A. Blockley, D. F. Walls, and H. Risken, Europhys. Lett.~\textbf{17}, 509 (1992).

\bibitem{Law96}C. K. Law and J. H. Eberly, Phys. Rev. Lett.~\textbf{76}, 1055 (1996).

\bibitem{Law98}B. Kneer and C. K. Law, Phys. Rev. A~\textbf{57}, 2096 (1998).

\bibitem{AminLTP} M. H. S. Amin, Low Temp. Phys.~\textbf{32}, 198 (2006).

\bibitem{Kippenberg}E. Verhagen, S. Deleglise, S. Weis, A. Schliesser, and T. J. Kippenberg, Nature~\textbf{482}, 63 (2012).

\bibitem{Esslinger}F. Brennecke, S. Ritter, T. Donner, and T. Esslinger, Science~\textbf{322}, 235 (2008).


\bibitem{Harris}J. D. Thompson, B. M. Zwickl, A. M. Jayich, F. Marquardt, S. M. Girvin, and J. G. E. Harris, Nature~\textbf{452}, 72 (2008).

\bibitem{Aspelmeyer}S. Groblacher, K. Hammerer, M. R. Vanner, and M. Aspelmeyer, Nature~\textbf{460}, 724 (2009).



\bibitem{Painter2}M. Eichenfield, R. Camacho, J. Chan, K. J. Vahala, and O. Painter, Nature~\textbf{459}, 550 (2009).


\bibitem{Lipson}G. S. Wiederhecker, S. Manipatruni, S. Lee, and M. Lipson, Opt. Exp.~\textbf{19}, 2782 (2011).

\bibitem{Carmichael} H. J. Carmichael, \textit{An Open Systems Approach to Quantum Optics}, (Springer-Verlag, Berlin, 1993).

\bibitem{Uhlmann} A. Uhlmann, Rep. Math. Phys.~\textbf{9}, 273 (1976).

\bibitem{Ian-opto}H. Ian, Z. R. Gong, Y. X. Liu, C. P. Sun, and F. Nori, Phys. Rev. A~\textbf{78}, 013824 (2008).


\bibitem{entanglephonon}X. W. Xu, Y. J. Zhao, and Y. X. Liu, Phys. Rev. A~\textbf{88}, 022325 (2013).


\end{thebibliography}
\end{document}